# Thermal expansion of solid solutions Kr-CH$_4$ at temperatures of liquid helium.

Aleksandrovskii A.N., Gavrilko V.G., Dolbin A.V., Esel'son V.B., Manzhelii V.G., Udovidchenko B.G.

B. Verkin Institute for Low Temperature Physics and Engineering of the National Academy of Sciences of Ukraine, Kharkov, Ukraine

*E-mail:* aalex@ilt.kharkov.ua

**Abstract.**

A negative contribution of the CH$_4$ impurity to the thermal expansion of the solution has been revealed in dilatometric studies of solid Kr+0.76% CH$_4$, Kr+5.25% CH$_4$ and Kr+10.5% CH$_4$ solutions at 1-23 K. It is shown that the negative contribution results from changes in the occupancy of the ground state of the A-modifications of isolated CH$_4$ molecules. Assuming that the CH$_4$ impurity singles and clusters contribute to the thermal expansion independently, we can estimate their contributions. The contribution of the singles to the thermal expansion of the solid solution is negative. The energies of the first excitational rotational states were determined for singles and two-body and three-body clusters of CH$_4$ molecules.

## 1. Introduction.

This is a study of the thermal expansion of solid Kr solutions containing 0.76, 5.2 and 10.5 mole % CH$_4$. A CH$_4$ molecule is a regular tetrahedron formed by hydrogen (protium) atoms with a carbon atom at the centre. Since the nuclear spin of carbon $^{12}$C is zero and the nuclear spin of the hydrogen atom is $I = \frac{1}{2}$, the rotational levels can have three types of symmetry corresponding to the total nuclear spin of methane. As a result, there are three modifications of the tetrahedral CH$_4$ molecules – A, T, E with total nuclear spin 2, 1, 0 respectively. For high-lying rotational levels the equilibrium concentrations of these nuclear spin CH$_4$ modifications are in the ratio 5:9:2. The A-modification has the lowest-energy rotational state. This state is five-fold degenerate and characterized by the rotational quantum number J=0. The levels having different symmetries relative to the nuclei do not combine much with each other [1]. Nevertheless, in experiment the spin modifications of CH$_4$ are observed to undergo mutual transformations (conversion). The transformation rate is to a great extent dependent on the condition experienced by the molecule. The conversion mechanism is not completely understood yet.

According to the theoretical calculation performed by Yamamoto, Kataoka and Okada [2, 3], in the high-symmetry crystal field the transition of CH$_4$ molecules into the rotational ground state (J=0) is accompanied by a reduction of their multipole part of the molecular interaction and thus leads to an increase in the crystal volume at lowering temperature (a negative thermal expansion coefficient). Let us denote this effect as the YKO (the first letters of the above authors' names) mechanism. It can readily be explained remembering that the rotational ground state (J=0) of a free CH$_4$ molecule is characterized by the spherically symmetrical probability distribution function, and thus the CH$_4$ molecule has a zero octupole moment in this state. In a crystal, the CH$_4$ molecule in this state more closely approaches the spherical shape than the molecules in other states and hence has a lower effective octupole moment.

At lowering temperature the growing occupancy of the rotational J=0 level can take contributions not only from the original A-modifications of $CH_4$ but from the E, T $\rightarrow$ A converted ones as well. The conversion is a much slower process than thermalization of the rotational spectra of the $CH_4$ modifications. The thermal expansion of the solution is determined by competition of two contributions – the phonon contribution responsible for the thermal expansion of the crystal lattice and the contribution of the rotational motion of the molecules. In the *fcc* case, the first contribution is positive, the other is negative (at sufficiently low temperatures). If the characteristic time of thermalization is considerably shorter than the characteristic time of conversion, the latter can be estimated from the time dependence of the sample length at fast-varying temperature. The effect of negative thermal expansion induced by the YKO mechanism [2,3] was observed experimentally only in solid $CH_4$ at helium temperatures [4-6].

The above consideration pertains both to disordered sublattices of solid $CH_4$ and some crystals including $CH_4$ as an impurity, to the solid $Kr+CH_4$ solution in particular. In the latter case the $CH_4$ molecule is in the octahedral crystal field of spherically symmetric Kr molecules.

If the above reasoning is correct, we may expect a negative contribution of the $CH_4$ impurity to the thermal expansion coefficient of the solid $Kr+CH_4$ solution at low temperatures. Using the dependence of the rotational energy spectrum of $CH_4$ upon the crystal field in the disordered phase, we can try to estimate the energy barriers impeding rotation of the $CH_4$ molecules in solid Kr. Besides, the time dependence of the solid $Kr+CH_4$ sample length at varying temperature of the sample surroundings was studied to derive information about the conversion rate of $CH_4$ molecules.

## 2. Experiment.

To answer the questions raised in this study, we have performed a dilatometric investigation of the thermal expansion of the solid Kr+0.76%, Kr+5.25% and Kr+10.5% $CH_4$ solutions at 1-23 K. At these temperatures, solubility of methane in krypton is about 80% [7]. The $CH_4$ concentration in the Kr+0.76% $CH_4$ solution is such that the contribution of isolated $CH_4$ molecules to the thermal expansion of the solid solution prevails. The Kr+5.25% $CH_4$ and Kr+10.5% $CH_4$ solution was intended to study the contributions of clusters of two or three $CH_4$ molecules to the thermal expansion. The coefficient of the linear thermal expansion of the solid $Kr+CH_4$ solutions was measured with a high-sensitivity capacitance dilatometer ($2 \cdot 10^{-9}$ cm resolution) [8]. The used krypton (99.937%) contained impurities of Ar (0.012%), $N_2$ (0.046 %) and $O_2$ (0.005 %). The purity of the gas was checked chromatographically. The methane used to prepare the solid $Kr+CH_4$ solution contained $N_2$ (0.03 %) and $O_2$ (<0.01 %). The solution was prepared at room temperature in a special measuring stainless steel vessel. The rated composition of the solution was controlled with a gas chromatographer. The solid samples were grown in the glass ampoule of the measuring cell of the dilatometer by condensation directly to the solid phase at ~ 63 K. The growth rate was about 2 mm/hour. The growth was controlled visually. The samples were transparent and without visible defects. The sample was then separated by thermal etching from the glass ampoule walls to make in the end a polycrystal cylinder about 22 mm in diameter, 24.6 mm (Kr+0.76 % $CH_4$), 31 mm (Kr+5.25 % $CH_4$) and 24.9 mm (Kr+10.5 % $CH_4$) high. The grain size was about 1 mm. The thermal expansion was measured along the cylinder axis. The dilatometer was also used to measure the thermal expansion of the matrix (solid Kr) grown by the same technique. The thermal expansion of the solutions was studied under isothermal conditions in the interval 1-23 K. Temperature was measured with germanium resistance thermometers. The distinctive feature of the experimental technique is that the sample temperature was changed by a jump of the power fed to the heater. Further on the power was kept constant. The time taken to fix the new temperature and the length of the sample varied from 0.5 to 1.5 hour. The temperature and length data were recorded every minute with the instruments, computer-processed in real time and displayed as plots and numerical data. The change-over to a new temperature was automatic and occurred after the sample attained such a "stable" state when its temperature changed not more than 0.01K during 10 minutes. The measurement was performed at lowering and rising temperatures.

### 3. Results and Discussion.

The contribution of the $CH_4$ impurity to the thermal expansion coefficient of the Kr- $CH_4$ solutions can be estimated when there is information about the thermal expansion of the matrix (pure solid Kr). To reduce the influence of experimental systematic errors, we performed our own investigation of the thermal expansion coefficient of pure Kr at 1-25K (this krypton was used in our solid solutions). Our coefficients and literature data [9] vary within 7% in the whole temperature range. The investigations of the thermal expansion coefficients of the solid Kr+0.76% $CH_4$, Kr+5.25% $CH_4$, Kr+10.5 % $CH_4$ solutions ($\alpha_{solution}$) were carried out. The $CH_4$ impurity brings the thermal expansion coefficient below that of pure solid Kr. The difference between the thermal expansion coefficients of the solid solutions and the Kr ($\alpha_{Kr}$) matrix is found as $\Delta\alpha = \alpha_{solution} - \alpha_{Kr}$.

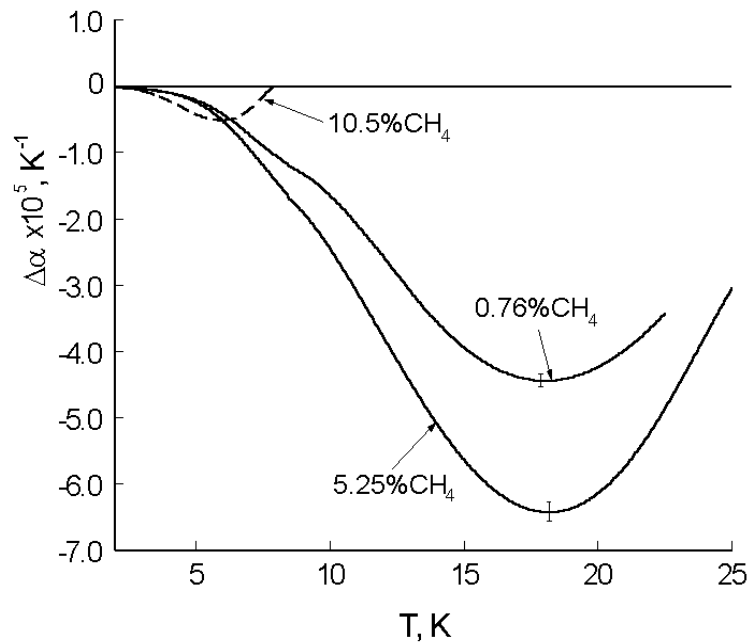

Fig.1. The contributions of the $CH_4$ impurity to the linear thermal expansion coefficient of solid solutions of Kr-$CH_4$.

The temperature dependence of $\Delta\alpha$ is shown in Fig. 1. It is interesting that $\Delta\alpha$ is negative for all the solutions in the whole temperature range. The result is however not surprising if we extend the thermal expansion calculation [2, 3] for solid $CH_4$ to our solutions, which is quite rightful. According to neutron diffraction data for a solid Kr+CH$_4$ solution [10, 11], the rotational spectrum of CH$_4$ in solid Kr is very close to that of orientationally disordered CH$_4$ molecules in solid CH$_4$. The dependence of the rotational spectrum of CH$_4$ molecules in the disordered sublattices of solid methane upon the coefficient f$_c$ characterising the crystal field value is shown in Fig. 2.

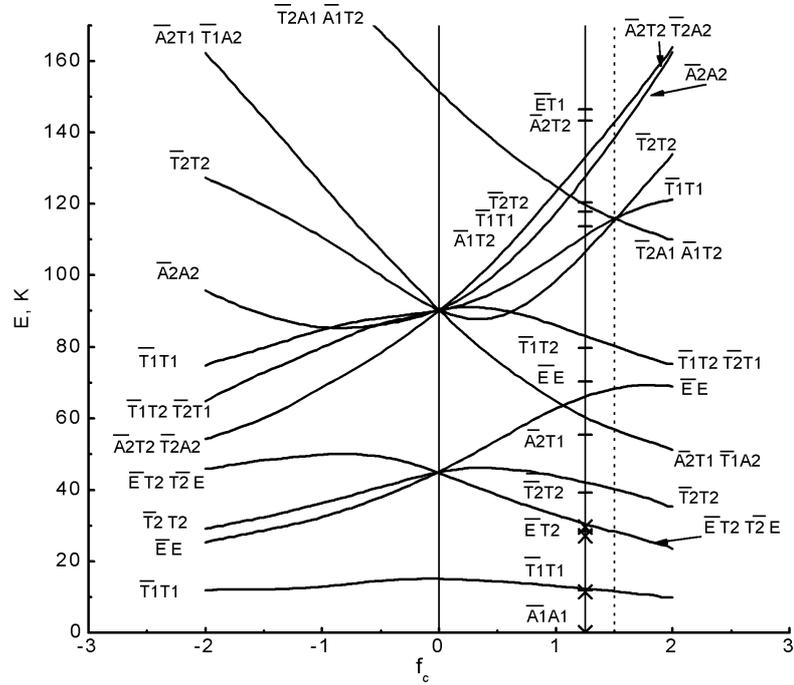

Fig. 2. The rotational spectrum of CH$_4$ molecules in a solid Kr-CH$_4$ solution and in solid pure CH$_4$ :
━ - Kr+CH$_4$ [13]; × - Kr+CH$_4$ [10];
solid lines - O$_h$ sublattice of antiferrorotational phase of solid CH$_4$ [3].

The authors of Refs.1, 2 believe that the crystal field value corresponding to the rotational spectrum section along the line f$_c$=1.25 (solid line in Fig. 2) is best matched to the experimental systematic of the rotational CH$_4$ levels in disordered sublattices.

The investigation of inelastic neutron scattering by the solid Kr+0.76% CH$_4$ solution has revealed the energy position of the peaks in the rotational transitions of CH$_4$ [10, 11]. The rotational energy levels [10] counted off from the ground state J=0 are shown in Fig. 2 (crosses). Unfortunately, at present no experimental data are available in literature for higher-energy part of the rotational spectrum of CH$_4$ in a Kr+CH$_4$ solution. Since the rotational spectra of CH$_4$ in a solid Kr+CH$_4$ solution and in the disordered phase are very close [2, 3], we will use the disordered-phase spectrum in our further calculation of the CH$_4$ behavior in a solid Kr+ CH$_4$ solution.

According to the above authors' opinion [2, 3], the negative thermal expansion of solid CH$_4$ follows from a decrease in the noncentral molecular interaction when the molecule changes into the ground-state. As a result, the crystal volume increases. The process is energy advantageous since the rotational ground state of the A-modification has the lowest energy. The excess thermal expansion Δα is thus dependent on the concentration of the A-modifications in the ground state. The temperature dependences of the equilibrium concentration of the A-modifications (solid line) and the equilibrium occupancy of its ground state (broken line) are shown in Fig.3. The curves were calculated for the equilibrium composition of spin modifications. The calculation was based on the rotational Kr+CH$_4$ spectrum obtained from experimental inelastic neutron scattering data for the states J=1, J=2. The values for the energy levels J=3 were taken from the rotational spectrum [2, 3]. As can be judged from the figure, the contribution of the YKO mechanism should manifest itself at temperatures of our experiments as occupancy of level J=0 appreciably changes at these temperatures. The conclusion agrees with the behaviour of the excess thermal expansion coefficient of the Kr+ 0.76% CH$_4$ and Kr+5.25% CH$_4$ solutions (see Fig. 1).

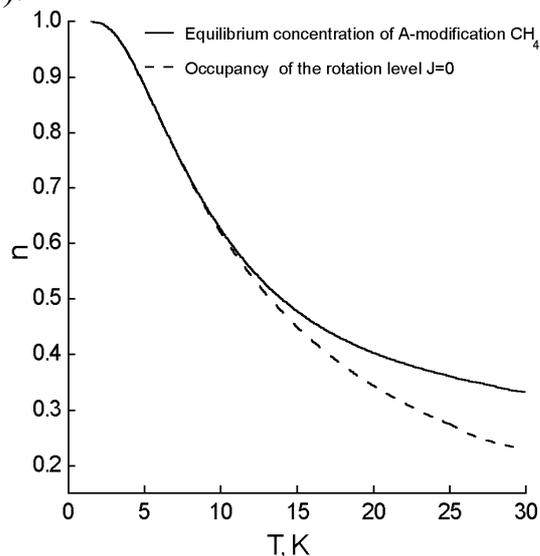

Fig. 3. Equilibrium concentration of the A-modification and the occupancy of its ground rotational state.

Apart from single impurity molecules, real solid solutions always contain their clusters. Assuming that the impurity singles and clusters contribute to the thermal expansion independently and assuming that the CH$_4$ molecules are distributed in the *fcc* lattice of Kr matrix randomly and using the method proposed in [12], we calculated the number of clusters consisting of one, two and three impurity molecules in solutions with 0.76, 5.25 and 10.5 %CH$_4$. The concentration of the CH$_4$ singles is then 0.69% in Kr in the solid Kr+0.76% CH$_4$ solution, 2.74% in Kr in the Kr+5.25% CH$_4$ solution, and 2.77% in Kr in the solution Kr+10.5% CH$_4$. Taking $\Delta\alpha$ of the solid solutions as a sum of the contributions from single, double and triple clusters and knowing their contents in the solution, we could estimate their specific contributions concerning to concentration *n* of singles, pairs and triples respectively (Fig. 4 and Fig. 4a). Contributions of various clusters to the thermal expansion of each solid solutions (a - Kr+0.76% CH$_4$, b - Kr+5.25% CH$_4$, c - Kr+10.5% CH$_4$) are showed on Fig. 5.

It should be noted, that before calculation we had to bring into correlation each value of temperature and the experimental value of the thermal expansion coefficient. For this purpose, the experimental values $\alpha$(T) of all the solutions were approximated by polynomial functions.

As our calculation shows, the main contribution to the negative thermal expansion of the solutions comes from the CH$_4$ singles. The pair clusters make a large positive contribution. The triple clusters produce only a slight effect on the thermal expansion of solid Kr-CH$_4$ solutions.

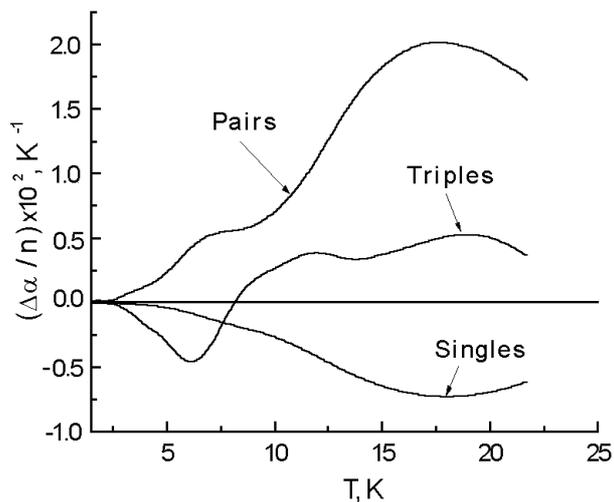

Fig.4. Specific contributions of various clusters to the thermal expansion of solid solution of Kr-CH$_4$ ($n$ - concentration of singles, pairs and triples respectively)

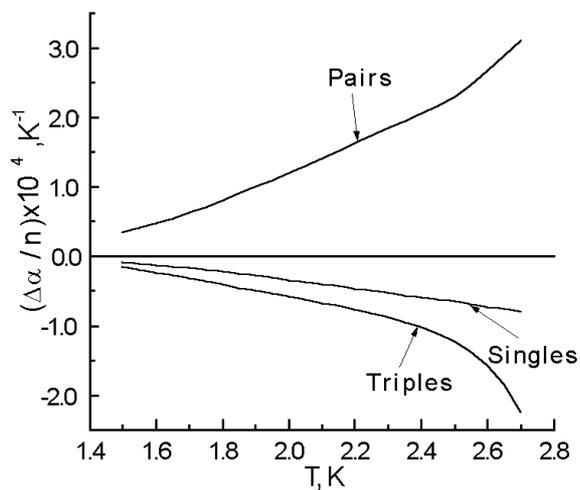

Fig.4a. Low temperature fragment of the specific contributions of various clusters to the thermal expansion of solid solution of Kr-CH$_4$ ($n$ - concentration of singles, pairs and triples respectively) .

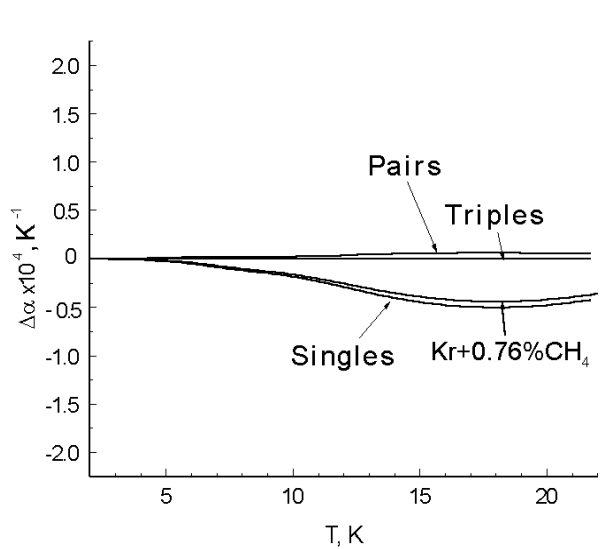

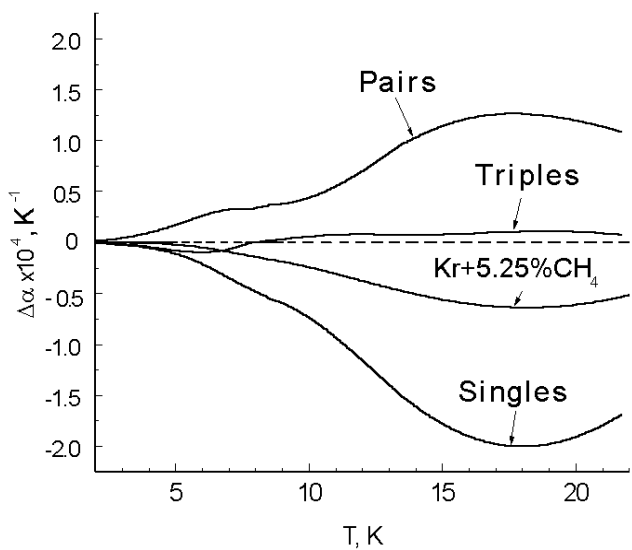

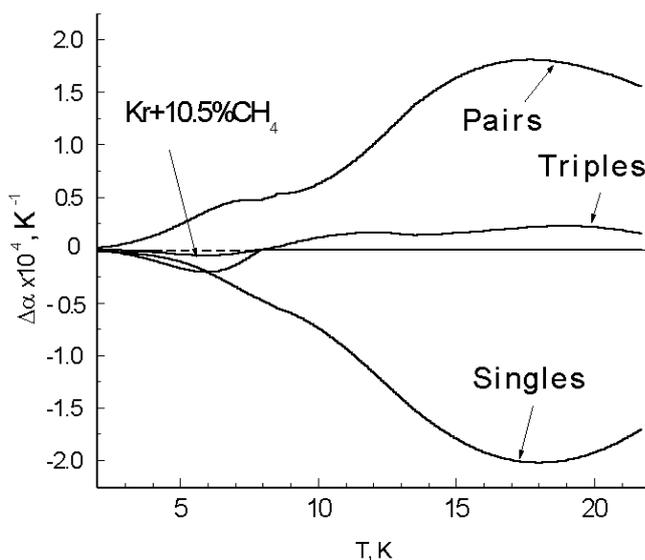

Fig. 5. Contributions of clusters consisting of 1, 2, and 3 impurity molecules of methane to thermal expansion of: a) Kr+0.76% CH$_4$ solution; b) Kr+5.25% CH$_4$; c) Kr+10.5% CH$_4$.

## 4. Contribution of matrix-isolated molecules CH₄ to thermal expansion of solid Kr-CH₄ solutions.

In this section we shall concentrate attention on consideration of behaviour of thermal expansion of Kr-CH₄ solutions at temperatures where the basic contribution to thermal expansion is brought by two lowest rotational levels.

If we extend the YKO mechanism [2, 3] explained the nature of negative thermal expansion of solid CH₄ to interpret the behavior of CH₄ impurity in a solid non-concentrated Kr-CH₄ solution, the negative contribution of impurity molecules to the thermal expansion of the solid Kr-CH₄ solution is no longer surprising. This extension is quite reasonable. According to neutron diffractometry [10, 11] data for solid Kr-CH₄ solutions, the rotational spectrum of CH₄ in solid Kr is close to that of the orientationally disordered CH₄ molecules in the antiferrorotational phase [3]. This fact is completely consistent with theoretical predictions [2, 13].

The authors [2, 3] believe that the crystal field value corresponding to the rotational spectrum cross-section along the line $f_c$=1.25 (the solid line in Fig.2) correlates most closely with experimental rotational CH₄ levels in the antiferrorotational phase. The energies of the peaks of the rotational transitions in CH₄ were found from inelastic neutron scattering data [10, 11] for the solid Kr+0.76% CH₄ solution. The rotational level energies of CH₄ counted off the ground state J=0 [10, 11] are shown in Fig. 2 (crosses). We can infer from the figure that the rotational CH₄ spectrum in the antiferrorotational phase and the rotational CH₄ spectrum in the Kr-CH₄ solution are fairly close, but in the solid Kr-CH₄ solution the section along the $f_c$=1.5 line provides better agreement of calculated and experimental results (the broken line in Fig.2). Proceeding from the similarity of the rotational CH₄ spectra in the solid Kr-CH₄ solution and in the antiferrorotational phase [2, 3], we shall use the spectrum in the antiferrorotational phase for our further calculation of the CH₄ behavior in the solid Kr-CH₄ solution. In [14] a special case is considered, when the free energy of the system can be written as a sum of free energies of the translational lattice vibrations and the rotational motion of the molecules. In this case the contribution of the rotational motion of noninteracting CH₄ molecules to the volume thermal expansion coefficient $\Delta\beta$ of the solid Kr-CH₄ solution can be presented as:

$$\Delta\beta = 3 \cdot \Delta\alpha = \frac{c\,\chi\,N_A}{VkT^2}\left\{\left\langle E^2\Gamma\right\rangle - \left\langle E\Gamma\right\rangle\left\langle E\right\rangle\right\} \tag{1}$$

Here $V$ is the molar volume, $\chi$ is the compressibility of the solution, $c$ – the molar concentration of the CH₄ in the solid solution Kr+CH₄ solution, $N_A$ – the Avogadro's number. The thermodynamic averaging is over the rotational CH₄ spectrum,

$$\left\langle ...\right\rangle = \sum_i (...) g_i e^{-E_i/kT} / \sum g_i e^{-E_i/kT}, \tag{2}$$

where $E_i$ are the rotational energy levels of CH₄; $g_i$ is their degeneracy multiplicity. $\Gamma_i = -\dfrac{\partial\ln E_i}{\partial\ln V}$ is the Grűneisen parameter of the $i$-th rotational level of CH₄.

Unfortunately, we cannot employ this expression for a quantitative comparison of calculated and experimental data on $\Delta\alpha$, since calculation involves contradictory $\Gamma_i$ values from different literature sources (e.g., see [3] and [15]). We therefore reduced our task and tried to estimate the energy gap between the ground and first excited rotational states of the CH$_4$ molecules from our experimental results. As the first step, we have found the temperature below which the contribution of the J>1 levels to $\Delta\alpha$ of the Kr+0.76% CH$_4$ solution is negligible. For that we calculated (see Eq.(1)) $\Delta\alpha$ for the case when only the two lowest levels A1A1 and T1T1 are involved in thermal expansion (Fig.2). The temperature dependence $\Delta\alpha$ calculated at the equilibrium distribution of the spin modifications of CH$_4$ is shown in Fig. 6. The $\Gamma_i$ values can be found from the rotational spectra of the orientationally disordered CH$_4$ molecules (spatial symmetry O$_h$) in the antiferrorotational phase of solid methane (Fig. 2).

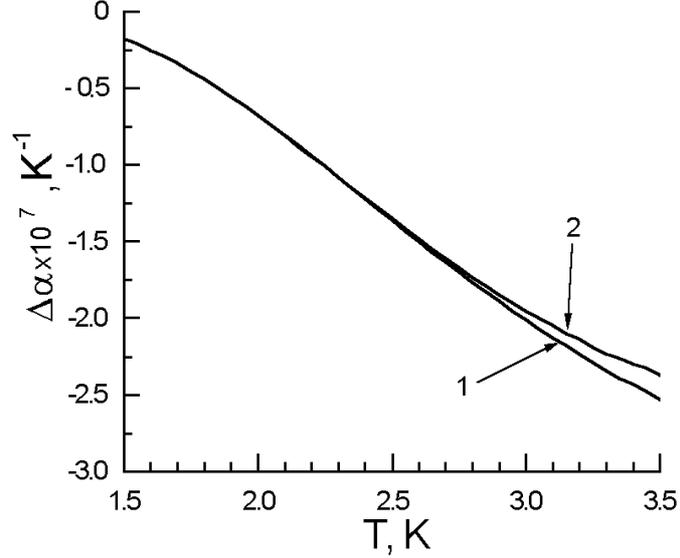

Fig.6. Low-temperature part of the calculated temperature dependence of the excess thermal expansion coefficient of the Kr+0.76% CH$_4$ solution:
1– for twelve low-lying rotational levels, 2 – for the two-level system.

Curve 1 is the $\Delta\alpha$ (T) involving twelve low-lying rotational levels (Fig. 2); curve 2 is calculated for the spectrum of only two lowest levels. It is seen that below 2.7 K both the thermal expansion coefficients coincide.

Using $\Delta E_{0-1}$ to denote the energy gap between the ground (A1A1) and the first excited (T1T1) states and $\Gamma_{0-1}$ for the Gruneisen coefficient calculated with respect to the ground state, Eq. (1) can be written as:

$$\Delta\alpha = \frac{c\,\chi\,N_A}{3VkT^2} \cdot \frac{g_1}{g_0} \cdot \Gamma_{0-1} \cdot \Delta E_{0-1}^2 \cdot e^{-\frac{\Delta E_{0-1}}{T}} \cdot \left(1 + \frac{g_1}{g_0} \cdot e^{-\frac{\Delta E_{0-1}}{T}}\right)^{-2} \approx$$

$$\approx \frac{c\,\chi\,N_A}{3VkT^2} \cdot \frac{g_1}{g_0} \cdot \Gamma_{0-1} \cdot \Delta E_{0-1}^2 \cdot e^{-\frac{\Delta E_{0-1}}{T}} \qquad (3),$$

where g$_0$ and g$_1$ are the degeneracies of the ground and the first excited states.

The sign of approximate equality in Eq. (3) is used because the factor within the brackets is equal approximately to unity. Our estimates show that at 1.5-2.7 K the bracketed factor varies from 1.00078 to 1.024. Since the isothermal compressibility $\chi$ and the molar volume $V$ are practically independent of temperature in this interval, in our calculation of $\Delta E_{0-1}$ we transformed Eq. (3) as

$$\ln\left(|\Delta\alpha| \cdot T^2\right) = B - \Delta E_{0-1} \cdot \frac{1}{T} \qquad (4),$$

where B is the temperature independent part. Equation (4) is a linear dependence on *1/T*, in which $\Delta E_{0-1}$ is the angular coefficient. Using Eq. (4) we found $\Delta E_{0-1}$ for the solid Kr+0.76% CH$_4$ solution. $\Delta E_{0-1}$ of this solution appears to be 11.23 K. The Kr+0.76% CH$_4$ was chosen due to the lowest contents of two-body and three-body clusters in it. Similarly, using Eq. (4) we found $\Delta E_{0-1}$ for the specific contributions of single molecules two-body and three-body clusters to the thermal expansion of our solutions (see Fig. 7).

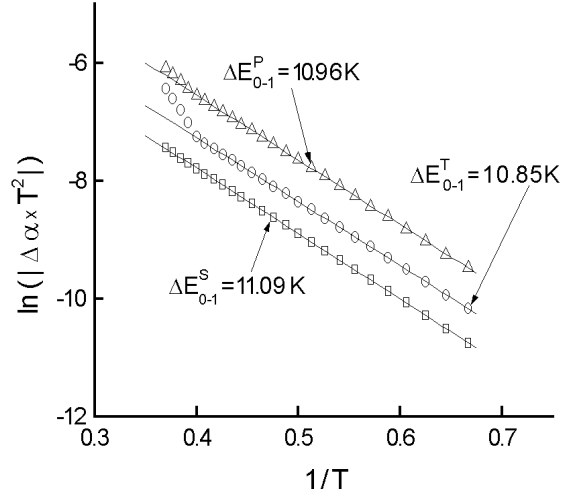

Fig. 7. Calculated dependences used to estimate $\Delta E_{0-1}$ in the libration spectra of single molecules (□), two-body (△), and three-body (○) clusters.

The $\Delta E_{0-1}$ values obtained for single molecules ($\Delta E_{0-1}^{S}$), two-body ($\Delta E_{0-1}^{P}$) and three-body ($\Delta E_{0-1}^{T}$) clusters are shown in Table 1. For comparison, the Table contains the corresponding energies calculated from inelastic neutron scattering date at T=4 K [11].

| Concentration of CH$_4$, mol % | $\Delta E_{0-1}^{S}$, K | $\Delta E_{0-1}^{P}$, K | $\Delta E_{0-1}^{T}$, K |
|---|---|---|---|
| 0.3 | 11.61 | - | - |
| 3.2 | 11.57 | 11.0[*] | - |
| 6.5 | 11.50 | 10.9[*] | 10.32[*] |
| This study | 11.09 | 10.96 | 10.85 |

• - data from the diagrams in [11].

Note that $\Delta E_{0-1}$ =11.7 K was obtained from studies of the heat capacity of the solid Kr+1% CH$_4$ +0.2% O$_2$ [16].

**Conclusions.**

Our dilatometric studies on solid Kr+CH$_4$ solutions show that the 0.76%, 5.25% and 10.5% CH$_4$ impurities make a negative contribution to the thermal expansion of the solution, the value of the contribution being nonproportionally dependent on the impurity concentration.

Assuming a random distribution of the CH$_4$ impurity over the *fcc* lattice of krypton, we were able to find the contributions of single, two-body and three-body clusters to the thermal expansion of the solutions. We have found that the negative contribution was mainly made by single molecules of the CH$_4$ impurity. As the CH$_4$ concentration increases, the negative contribution of the single molecules is partially balanced by the positive contribution of the two-body clusters. In low-concentrated solid Kr-CH$_4$ solutions the thermal expansion is thus determined by the competition of positive contributions made by the thermal expansion of the matrix lattice and two-body clusters and, on the other hand, of the negative contribution of the single impurity molecules.

Using our experimental results on the thermal expansion of a solid solution with the highest concentration of single impurity molecules (Kr+0.76% CH$_4$), we could estimate the energy gap between the ground and first excited rotational states. Also, $\Delta E_{0-1}$ was found for single impurity molecules, two-body and three-body clusters. The results obtained are compared with literature data.